\documentclass[conference]{IEEEtran}
\IEEEoverridecommandlockouts
\usepackage{cite}
\usepackage{amsmath,amssymb,amsfonts}
\usepackage{algorithmic}
\usepackage{graphicx}
\usepackage{textcomp}
\usepackage{xcolor}
\usepackage{threeparttable}
\usepackage{dblfloatfix}
\usepackage{adjustbox}
\usepackage{caption}
\usepackage{subcaption}
\usepackage{multirow}
\def\BibTeX{{\rm B\kern-.05em{\sc i\kern-.025em b}\kern-.08em
    T\kern-.1667em\lower.7ex\hbox{E}\kern-.125emX}}
\usepackage{comment}
\usepackage{authblk}

\def\BibTeX{{\rm B\kern-.05em{\sc i\kern-.025em b}\kern-.08em
    T\kern-.1667em\lower.7ex\hbox{E}\kern-.125emX}}

\usepackage{fancyhdr}
\fancypagestyle{firstpage}
{
    \fancyhead[L]{\footnotesize © 2022 IEEE.  Personal use of this material is permitted.  Permission from IEEE must be obtained for all other uses, in any current or future media, including reprinting/republishing this material for advertising or promotional purposes, creating new collective works, for resale or redistribution to servers or lists, or reuse of any copyrighted component of this work in other works. This paper is accepted at the 28th IEEE International Symposium on On-Line Testing and Robust System Design (IOLTS) 2022.}
    \fancyhead[R]{}
}

\begin{document}

\title{A Novel Fault-Tolerant Logic Style with Self-Checking Capability}
\author[1]{Mahdi Taheri}
\author[2]{Saeideh Sheikhpour}
\author[3]{Ali Mahani}
\author[1]{Maksim Jenihhin}
\affil[1]{Tallinn University of Technology, Tallinn, Estonia}
\affil[2]{Ghent University, Ghent, Belgium}
\affil[3]{Shahid Bahonar University of Kerman, Kerman, Iran}
\affil[1]{mahdi.taheri@taltech.ee}

\IEEEoverridecommandlockouts \IEEEpubid{\makebox[\columnwidth]{978-1-6654-7355-2/22/\$31.00~\copyright2022 IEEE \hfill} \hspace{\columnsep}\makebox[\columnwidth]{ }}
\maketitle
\thispagestyle{firstpage}
 \IEEEpubidadjcol
\begin{abstract}
We introduce a novel logic style with self-checking capability to enhance hardware reliability at logic level. The proposed logic cells have two-rail inputs/outputs, and the functionality for each rail of outputs enables construction of fault-tolerant configurable circuits. The AND and OR gates consist of 8 transistors based on CNFET technology, while the proposed XOR gate benefits from both CNFET and low-power MGDI technologies in its transistor arrangement. To demonstrate the feasibility of our new logic gates, we used an AES S-box implementation as the use case. The extensive simulation results using HSPICE indicate that the case-study circuit using on proposed gates has superior speed and power consumption compared to  other implementations with error-detection capability.
\end{abstract}


\section{Introduction}
%Advanced Encryption Standard (AES) is selected as the secret key encryption standard by the National Institute of Standards and Technology (NIST) in 1997 \cite{aes}. Until now, AES is widely adopted for a variety of applications to encrypt sensitive data. The recent wide adoption of applications such as internet of things (IoT), wireless sensor networks and RFID tags which have had an undeniable effect on our daily life, has emphasized the demand for hardware security. These applications are inherently low-cost with tight constraints on their area and power budgets. Thus, it is vital to optimize hardware implementation of cryptographic primitives in terms of power consumption and implementation area. 
The advancements in circuit design technology, such as increased routing complexity and operating clock frequency, the increased integration density of transistors, along with the decreasing transistor size and power supply voltage, are important reasons leading to growing fault rates in the integrated circuits (ICs). Along with the increasing demand for fault-tolerance against unintentional faults during the ICs' lifetime caused by reliability issues, fault tolerance against malicious faults is also getting more prominent. In the implementation of cryptographic primitives and embedded systems, security attacks by controlled fault injection have a purpose to extract secret information (e.g. \cite{aes, Lai-vlsi-soc20}) and here the fault tolerance aspect becomes may even more crucial.

Due to the presence of random and malicious faults, reliability has become a crucial concern for the hardware implementation\cite{sbox0}. Thus, it is essential to develop innovative countermeasures to guarantee the reliability and therefore, security of confidential data  \cite{repo}.
In the crypto-accelerators hardware domain, the current countermeasures proposed for this purpose is categorized into two categories, i.e. detection and infection \cite{repo}. Faults are detected in the execution time of the device, in the detection countermeasures. The output (cipher-texts) is not generated when a fault is detected. This action prevents potential exploitation in fault attacks. The logical effect of faults is changed so that the cipher-text is affected in a different way than the attacker expects, in infection countermeasures. As a result, attackers cannot further analyze faulty cipher-texts. 

\begin{figure}[h]
    \centering
    \includegraphics[width=0.5\textwidth]{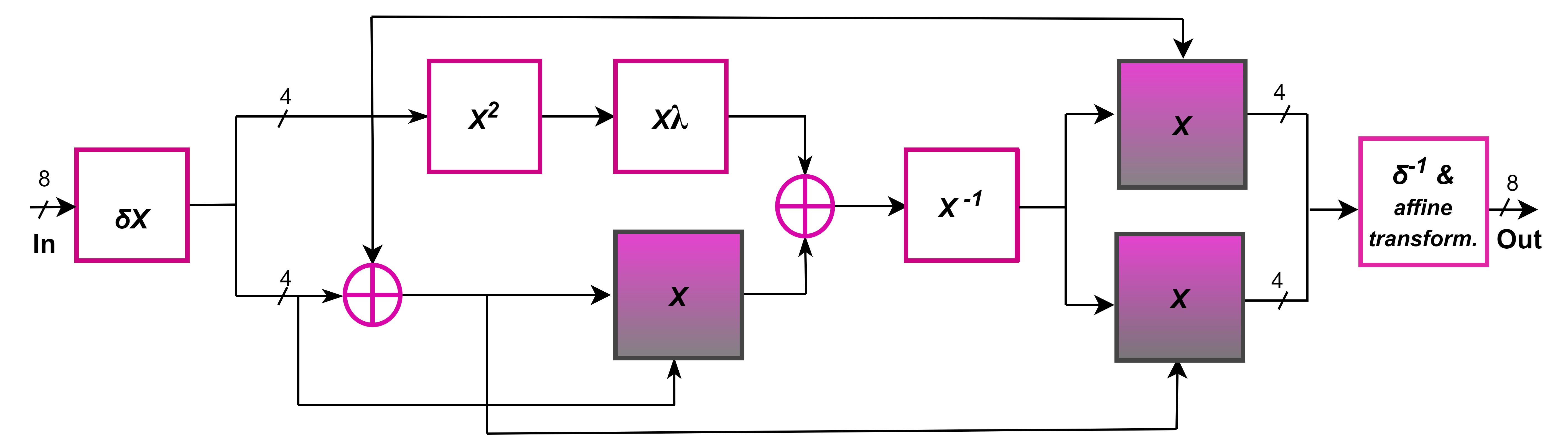}
    \caption{Case-study design architecture: Composite Field based S-box.}
    \label{fig:sbox}
\end{figure}

In this paper to reduce the fault effects, we present a detection countermeasure at the transistor level. In particular, we introduce a low-cost self-checking logic style including AND, XOR and OR gates for reliable implementation of any digital circuit. 
%, specifically cryptographic algorithms implementation as the use case like \cite{sbox0}. 
The transistors are organized in a structure in our proposed gates such that they can detect any single stuck-at-fault on their inputs and also at each transistor.
The area and power-consumption of any circuit implementation are essential parameters for its functional and extra-functional aspects such as security \cite{sheikhpur2021strengthened}.
In this paper, we focus on the area-efficient low-power and implementations of circuits using our proposed logic style. As a case-study we analyze the effect on a AES S-box design. There exist many reported implementations for the S-box of AES considering varying design metrics such as power, area and delay for various applications \cite{sbox1,sbox2,sbox3,sbox5,sbox6,sbox7,sbox8, taheri2020dmr}. Without loss of generality, the composite field based S-box in \cite{sbox1} is selected as a case study in this paper (see Fig. \ref{fig:sbox}).
\\
Our main contributions are as follows:
%\vspace{-0.1in}
\begin{itemize}
\item We introduce a novel low-cost logic style with self-checking capability for digital circuit secure implementation, especially aiming at cryptographic algorithms.
\item We prove that the proposed cells detect 100$\%$ of single-bit faults
\item We implement a case-study design (an AES S-box core) using the proposed logic cells.
\item We validate the feasibility and efficiency of our proposed  logic style on the case-study circuit for variation of the power supply and output capacitance using HSPICE. 
\item Finally, we simulate several circuits with fault detection capability with the same situation with our proposed circuit to have a fair comparison.   
\end{itemize}

\section{Related Work}
A US government standard for security FIPS 140, emphasizes that the physical security mechanisms must provide complete protection for the cryptographic device with the purpose of detecting and responding to the unauthorized attempts at physical access \cite{fips}.  Among such mechanisms are detection techniques that can be classified into hardware redundancy, time redundancy, information redundancy, hybrid redundancy and self-checking techniques.
Dual Modular Redundancy (DMR) is a well-known configuration of hardware redundancy for detection. This technique offers a high level of reliability in an almost 100 $\%$ hardware overhead, as it duplicates the hardware unit \cite{dmr-ref, dmr, taheri2021fault}.

Time redundancy can be achieved by reusing the same hardware unit, by applying the same input and then performing a comparison between the outputs one execution to the other for fault detection purpose \cite{time}.
Double Time Redundancy (DTR) is a simple detection countermeasure based on time redundancy. %\cite{dtr}.
DTR technique greatly reduces the hardware overhead, but introduces high time penalty and consequently more than 100 $\%$ performance degradation. Another drawback of this technique is its inability to detect permanent faults. \\
The well-known form of information redundancy is error detection coding \cite{time}. In these techniques, several redundant bits are generated from the input blocks; then these redundant bits propagate along with their correspondent input blocks and are checked when the output block is generated.
In \cite{sbox0}, a low-cost parity-based error detection technique is proposed for the composite field-based S-box of AES. The authors divide the S-box into five blocks and introduce a parity bit generator circuit for each of those blocks.

The self-checking technique is an efficient detection countermeasure which significantly reduces the area and power overheads associated with hardware redundancy, and puts little additional throughput and performance degradation as compared with the time redundancy techniques. This type of countermeasure also offers a higher level of reliability and in consequence security than error information redundancy based detection techniques. A self-checking technique in the transistor level for AES S-box is proposed in\cite{self}. In fact, the authors in \cite{self} propose logic gates using the Pseudo-nMOS gate structures with the capability of self-checking. This logic style suffers from a significantly higher power consumption because of static current in Pseudo-nMOS gate structures \cite{nmos}.

\section{Proposed Self-Checking Logic Style}
%The AES algorithm contains four main operations: SubBytes, ShiftRows, MixColumns and AddRoundKey using a round function. The SubByte is the only non-linear operation of AES which is applied independently on each byte of the intermediate state matrix using 16 substitution boxes (S-boxes). It is the most area-consuming operation whose non-linearity is used for protecting AES against linear cryptOanalysis \cite{sbox1,sboxm}.
In this paper, we propose a full-swing logic style with self-checking capability that is Power-Delay-Product efficient and thus called \emph{Self-Checking Power-Delay-Product (SCPDP)}.
%The proposed logic style provide a secure and reliable implementation for cryptographic primitives, here, the S-box of AES as a case study.

Let $s$ and $\bar{s}$ be the inputs to a self-checking unit $f$ and $f(s)$ and $\bar{f(s)}$ be its outputs. If $f$ is a proper and well-designed self-checking unit, then a fault in this unit will affect $f(s)$ and $\bar{f(s)}$ in a different way. In consequence, the two outputs $f(s)$ and $\bar{\bar{f(s)}}$ will not match and so faults are detected.\\ 
Fig. 2 show the proposed structure of our proposed cells. This logic style, includes AND, OR and XOR gates as basic logic gates by which all digital circuits can be constructed. As shown in Fig. 2, all of the proposed components have two-rail inputs and also two-rail outputs. In fact, each gate is fed by an input (e.g. $A$) and its complement (e.g. $\bar{A}$), as a two-rail input, and it produces the corresponding output (e.g. $O$) and its complement (e.g. $\bar{O}$), as a two-rail output.

It should be noted that our proposed logic style do not need to invert logic. The inverting operation is performed simply by swapping each output's rails.
The $SCPDP$'s logic cells in Fig. 2 are based on low-power MGDI (Modified Gate Diffusion Input)  technique \cite{mgdi} and CNFET (Carbon Nanotube Field-Effect Transistor) technologies. It is worth mentioning that "CNTFET uses much less power than a silicon-based device and have high speeds"\cite{obite2019carbon}. Note, all components belonging to this logic style are identical except for the inputs and are suitable for constructing reliable and secure configurable platforms. Each self-checking component of this family consists of 8 transistors.\\

\begin{figure*}
\centering
\subcaptionbox{AND-SCPDP}{\includegraphics[width=0.33\textwidth]{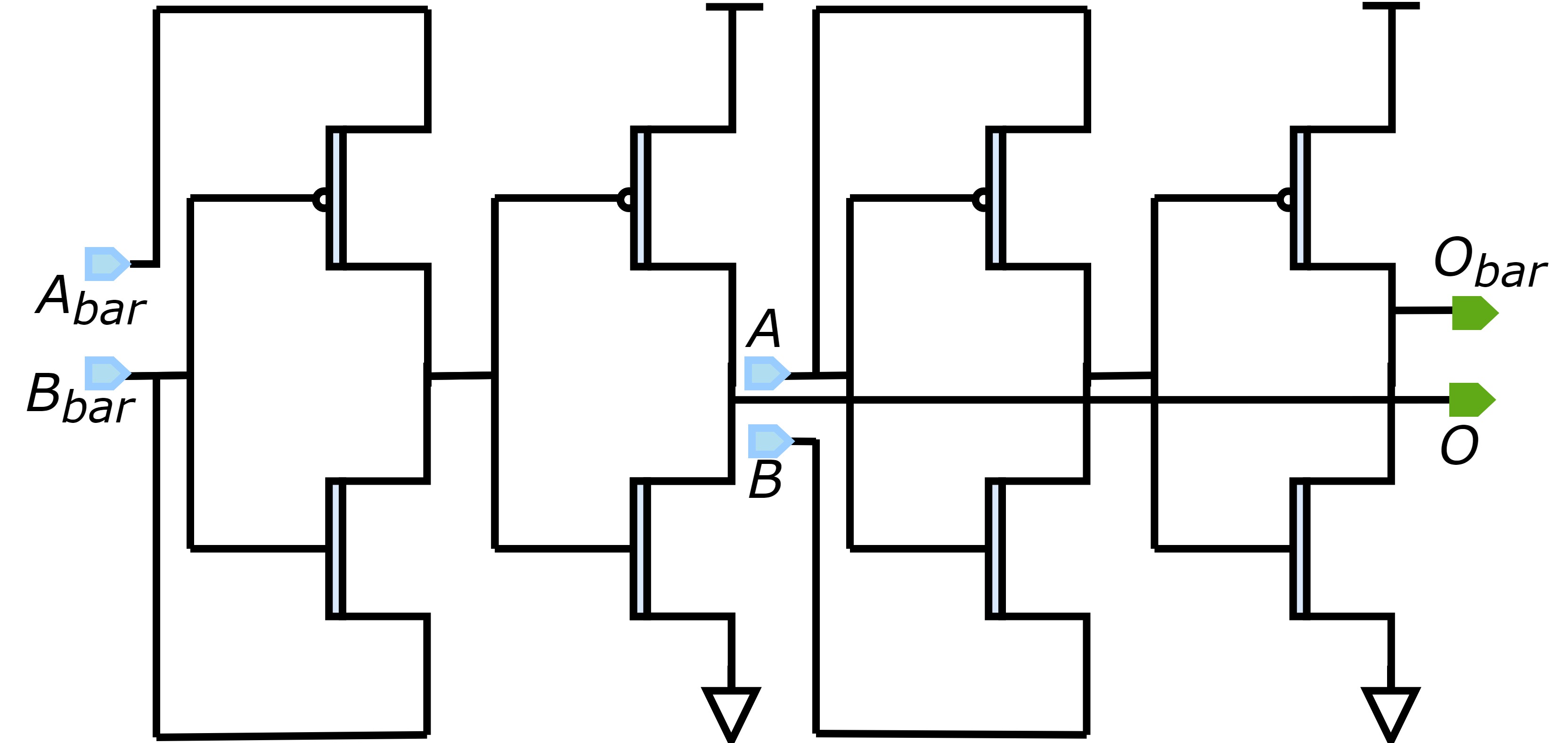}}%
\hfill
\subcaptionbox{OR-SCPDP }{\includegraphics[width=0.33\textwidth]{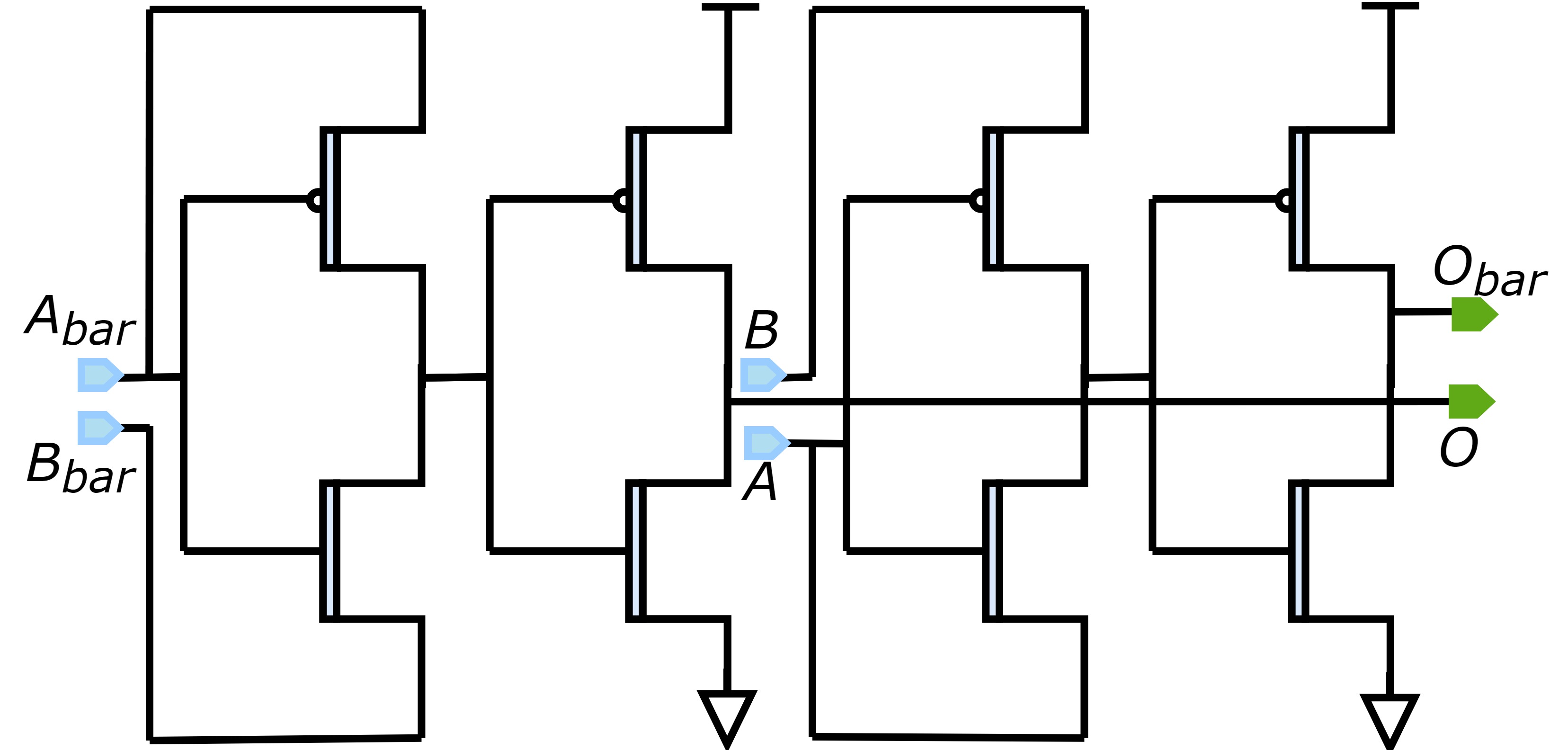}}%
\hfill
\subcaptionbox{XOR-SCPDP }{\includegraphics[width=0.33\textwidth]{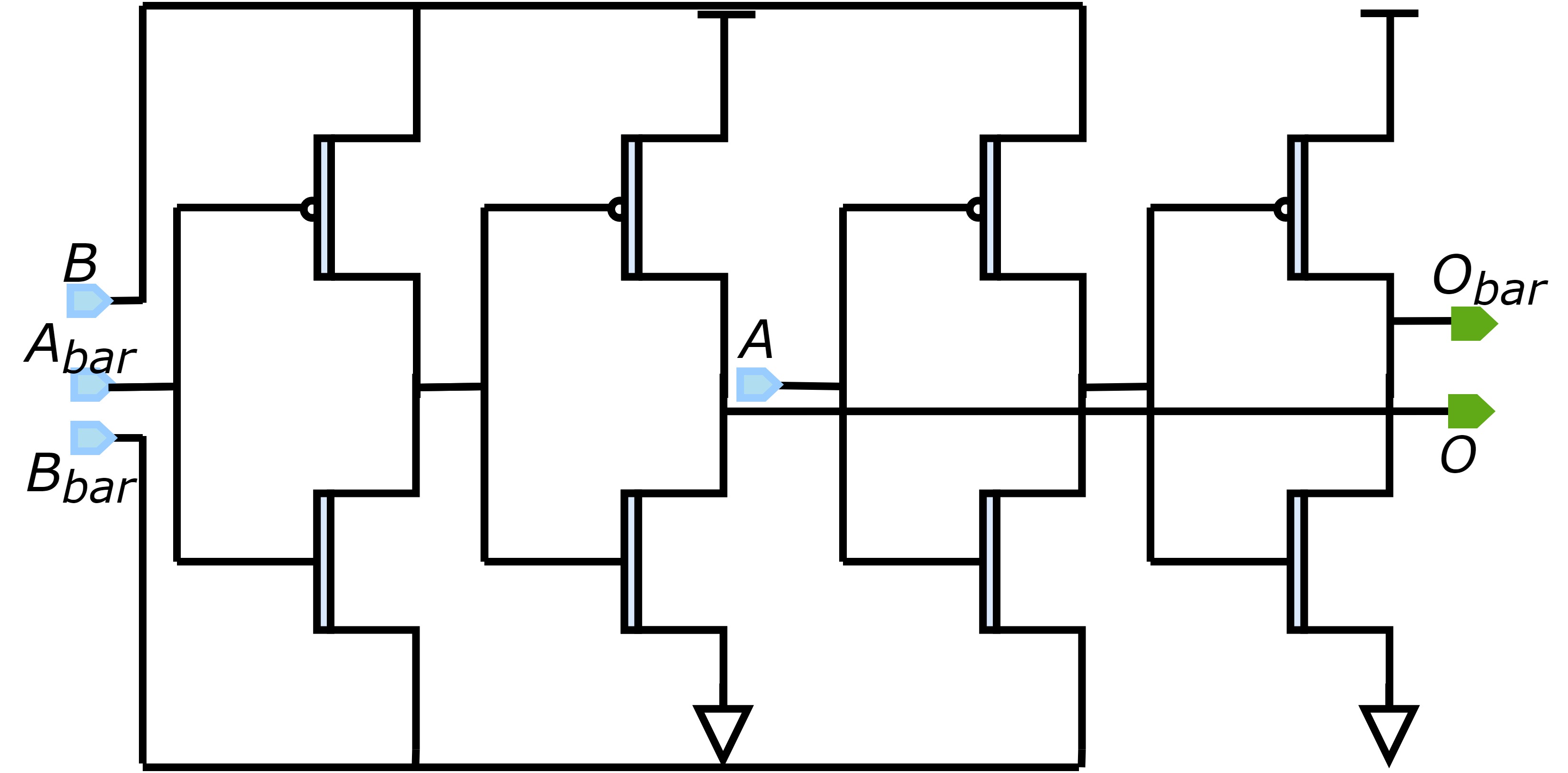}}%
\hfill
%\subcaptionbox{AND-SCA }{\includegraphics[width=0.25\textwidth]{and-nand2.jpg}}%
%\hfill
%\subcaptionbox{OR-SCA }{\includegraphics[width=0.25\textwidth]{or-nor2.jpg}}%
%\hfill
%\subcaptionbox{XOR-SCA }{\includegraphics[width=0.43\textwidth, scale=2]{xor-xnor2.jpg}}%
\caption{Proposed Gate structures with Self-Checking capability and Power-Delay-Product efficiency (SCPDP).}
\label{fig:and1}
\end{figure*}
In the presence of any type of single stuck-at faults, each component produces a non-valid output (11 or 00) according to Fig. 2. Again, each component produces a non-valid output when it takes a non-valid value (11 or 00) as its input.\\
Later, in the next subsection, we prove that the case of cells having single stuck-at faults in conjunction with non-valid inputs will be considered, in order to show the propagation of faults to the output and the likelihood of detecting multiple faults. Eq. \ref{eq:and1o}-6 show the output signal equations for each designed component.\\

AND-SCPDP gate (Fig 2a):
\begin{equation}
\begin{split} 
\label{eq:and1o}
O &=\overline{\overline{B_{bar}}.A_{bar} + B_{bar}.B_{bar}} = (B_{bar}+\overline{A_{bar}}).\overline{B_{bar}}\\& =  \overline{B_{bar}} . \overline{A_{bar}}
\end{split} 
\end{equation}
\vspace{-0.1in}
\begin{equation}
\label{eq:and1obar}
O_{bar}=\overline{\overline{A}.A + A . B} = \overline{A . B} 
\end{equation}

OR-SCPDP gate (Fig 2b):
%\vspace{-0.1in}
\begin{equation}
\label{eq:or1o}
O=\overline{\overline{A_{bar}}.A_{bar} + A_{bar} . B_{bar}} = \overline{A_{bar}} + \overline{B_{bar}} 
\end{equation}
\vspace{-0.2in}
\begin{equation}
\label{eq:or1obar}
O_{bar}=\overline{\overline{A}.B + A .A} = (A +\overline{B}).\overline{A} =  \overline{B} . \overline{A} = \overline{A + B}
\end{equation}

XOR-SCPDP gate (Fig 2c):
\begin{equation}
\label{eq:xor1o}
\begin{split}
O & =\overline{\overline{A_{bar}}.B + A_{bar} . B_{bar}} = (A_{bar} + \overline{B}) . (\overline{A_{bar}} + \overline{B_{bar}}) \\&  =
A_{bar} . \overline{B_{bar}}  +  \overline{B} . \overline{A_{bar}}
\end{split}
\end{equation}
\vspace{-0.1in}
\begin{equation}
\label{eq:xor1obar}
O_{bar}=\overline{\overline{A}.B + A . B_{bar}} = (A + \overline{B}) . (\overline{A} + \overline{B_{bar}}) = A . \overline{B_{bar}}  +  \overline{B} . \overline{A}
\end{equation}

\section{Specification of the proposed gates}
Our proposed logic gates detect all single-bit faults at inputs or some intermediate signals due to their structure,
which means \textbf{if a single-bit fault affects the input signal of the proposed gates, it will be detected.} It should be noted that if there is no error, $\overline{B_{bar}}$, $\overline{A_{bar}}$ are equal to $B$ and $A$, respectively.

We use Boolean Difference (BD) method to prove our claim. It is used to analyze the effect of errors on the outputs of combinational logic circuits \cite{bd}. For this purpose, BD of each gate's outputs, i.e. $O$ and $O_{bar}$, with respect to different input signals, are calculated.

Generally the BD which indicates the dependency an output $f$ on the signal signals, e.g. $s$, is defined as follows:
\begin{equation}
\frac{df}{ds} = f(s=0) \oplus f(s=1)
\end{equation}
when $\frac{df}{ds} \neq 1$, change in s cannot affect $f$.\\
Here, we prove the gates dependencies for just for the XOR gate (Eq.(\ref{eq:xor1o}, \ref{eq:xor1obar})). For other gates the same method can be applied.

The $Self-checking$ XOR Gate:\\
We check the dependency of the proposed XOR outputs on its input signals as follows:

To find out the dependency of $SCPDP$'s XOR Gate output signal $O$ (Eq. \ref{eq:xor1o}) on its input signals, we calculate the BD as:

\begin{equation}
\label{eq:xora}
\frac{dO}{dA} = (A_{bar} . \overline{B_{bar}}  +  \overline{B} . \overline{A_{bar}}) \oplus (A_{bar} . \overline{B_{bar}}  +  \overline{B} . \overline{A_{bar}}) = 0
\end{equation}

\vspace{-0.2in}

\begin{flalign}
\label{eq:xorb}
\begin{split}
\frac{dO}{dB} & = (A_{bar} . \overline{B_{bar}}  +  \overline{0} . \overline{A_{bar}}) \oplus (A_{bar} . \overline{B_{bar}}  +  \overline{1} . \overline{A_{bar}} ) \\& = \overline{A_{bar}}
\end{split} 
\end{flalign}

\vspace{-0.2in}
\begin{equation}
\label{eq:xorabar}
\frac{dO}{dA_{bar}} = (0 . \overline{B_{bar}}  +  \overline{B} . \overline{0}) \oplus (1 . \overline{B_{bar}}  +  \overline{B} . \overline{1}) = B_{bar} \oplus \overline{B_{bar}}
\end{equation}
\vspace{-0.2in}
\begin{equation}
\label{eq:xorbbar}
\frac{dO}{dB_{bar}} = (A_{bar} . \overline{0}  +  \overline{B} . \overline{A_{bar}}) \oplus (A_{bar} . \overline{1}  +  \overline{B} . \overline{A_{bar}}) = A_{bar}
\end{equation}

Similarly, according to Eq. \ref{eq:xor1obar}, the BD of output signal $O_{bar}$ with respect to input signals can be
calculated as: 

\begin{equation}
\label{eq:xor2a}
\frac{dO_{bar}}{dA} = (0 . \overline{B_{bar}}  +  \overline{B} . \overline{0}) \oplus (1 . \overline{B_{bar}}  +  \overline{B} . \overline{1}) =  \overline{B} \oplus \overline{B_{bar}}
\end{equation}
\vspace{-0.15in}
\begin{equation}
\label{eq:xor2b}
\frac{dO_{bar}}{dB} = (A . \overline{B_{bar}}  +  \overline{0} . \overline{A}) \oplus (A . \overline{B_{bar}}  +  \overline{1} . \overline{A}) = \overline{A}
\end{equation}
\vspace{-0.15in}
\begin{equation}
\label{eq:xor2abar}
\frac{dO_{bar}}{dA_{bar}} = (A . \overline{B_{bar}}  +  \overline{B} . \overline{A}) \oplus (A . \overline{B_{bar}}  +  \overline{B} . \overline{A}) = 0
\end{equation}
\vspace{-0.15in}
\begin{equation}
\label{eq:xor2bbar}
\frac{dO_{bar}}{dB_{bar}} = (A . \overline{0}  +  \overline{B} . \overline{A}) \oplus (A . \overline{1}  +  \overline{B} . \overline{A}) = A
\end{equation} 
Eq. \ref{eq:xora} and \ref{eq:xor2abar} show that the output signals $O$ and $O_{bar}$ don't depend on the signals $A$ and $A_{bar}$, respectively, because $\frac{dO}{dA}=0$ and $\frac{dO_{bar}}{dA_{bar}}=0$. Therefore, the fault on these signals will definitely be detected. The dependency of $O$ and $O_{bar}$ on input signal $B$ are given in Eq. \ref{eq:xorb} and \ref{eq:xor2b}, respectively. These equations state that  $\frac{dO}{dB}=\overline{A_{bar}}$ and $\frac{dO_{bar}}{dB}=\overline{A}$. Here, as the fault assumed to be on signal $B$, and we allow only single-bit faults, the other signals are fault-free. The $\overline{A_{bar}}$ and $\overline{A}$ must carry different values when are fault-free, i.e. $\overline{A_{bar}}=0$ and $\overline{A}=1$ or $\overline{A_{bar}}=1$ and $\overline{A}=0$. Therefore, the fault on input signal $B$ will be detected. 
The fault at input signal $B_{bar}$ changes simultaneously both outputs if $A_{bar} = A =1$, See Eq. \ref{eq:xorbbar} and \ref{eq:xor2bbar}. As $B_{bar}$ is faulty, $A$ and $A_{bar}$ must be fault-free in single-fault assumption and must carry different logical values. As a result, fault on $B_{bar}$ can not fail our detection mechanism.
To find out the effect of faults at intermediate connections, we can calculate the BD of the output signals with respect to the intermediate signals and proceed similarly. \\
Hence, the single-bit faults are either functionally masked or detected by the proposed logic style. 

Due to the presence of convergent paths in a circuit, a single-bit error may appear at an output of several gates. It is worth mentioning that the our proposed gates detect the fault in the case if the common signal enters the first or second input of XOR gate. To check how our logic style detects faults in this situation, we must similarly calculate the functionality of that gate and then find its dependency on all input pairs. The BD in these cases verify that the proposed gates can detect faults.

As discussed, our logic style can tolerate all single-bit faults. While providing fault resistance against multiple faults is much more interesting. We perform fault injection in simulation level, for 40,000,000 stuck-at-0 and stuck-at-1 multiple random fault models separately, with random plain-text and input key at each simulation of our case-study design for both burst and random fault models. 

It is worth mentioning that for each fault injection simulation the random round inputs and random locations are generated using a random number generator. 
%Given that undetected faulty cipher-texts can be exploited to extract the secret information in fault attack-based crypto-analysis, the 
In the proposed approach, Fault Coverage (FC) is an essential fault tolerance metric, which is defined as follows:

\begin{equation}
\label{eq:fc}
FC=\frac{\#\text{ of all faults}- \#\text{ of undetected faults}}{\#\text{ of all faults}}
\end{equation}

The FC results for multiple-bit fault injection of various sizes; i.e. 1-bit to 14-bit fault sizes; are shown in Table. \ref{tab:fault_injection} for $SCPDP$. As shown in this Table, FC for the proposed logic style is 99.98 $\%$ for random stuck-at fault sizes. The reported simulation results in Table. \ref{tab:fault_injection}  show that our proposed gates are not able to detect only a small fraction of multiple-bit faults. Thus, the proposed logic style provides high fault tolerance against multiple-bit faults, but also, in the case of the case-study design, high security against malicious fault injection attacks. It is worth mentioning that the fault coverage of the proposed approach is significantly higher than in the other compared approaches e.g. 97$\%$ in \cite{sbox0}.

\begin{table}[h]
\centering
\caption{Fault injection for 40,000,000 stuck-at-0 and stuck-at-1 multiple ran-
dom fault models.}
\label{tab:fault_injection}
\begin{tabular}{|c|c|c|c|c|}
\hline
Fault coverage &
  \begin{tabular}[c]{@{}c@{}}S-a-0 \\  fault in SCPDP\end{tabular} &
  \begin{tabular}[c]{@{}c@{}}S-a-1\\   fault in SCPDP\end{tabular} 
  \\ \hline
Burst faults &
  99.831\% &
  99.83\% 
 \\ \hline
Random faults &
  99.98\% &
  99.98\% 
\\ \hline
\end{tabular}
\end{table}

\section{Evaluation and experimental results}
Based on the previous discussions, in this section, we evaluate the synthesized results on a practical case-study AES S-box design implementation by applying our proposed logic style and the state-of-the-art gates (using the 32nm CNFET technology with HSPICE Synopsys tool). All the circuits have been simulated using through this process technology and were supplied with two 1.2 V and 0.8 V with three different output capacitance. We compare our results with the other works under six different states called as Test Bench 1-6 (TB1 to TB6) in which the specification of each TB can be found in Table \ref{tbl:t1}. It is important that the reference methods are base models and are widely used in recent works, e.g. \cite{gaded2019composite, bintalib2021hybrid, seyedi2022nwise}. For our simulations, we have chosen the size of transistors in a way that the minimum PDP is achieved for the circuit.
We consider the compact S-box based on the polynomial basis, using composite-field arithmetic as our circuit under test.
\begin{table}[h]
\centering
\caption{Different combination of input voltage and output capacitance.}
\label{tbl:t1}
\begin{tabular}{|l|l|r|}
\hline
\multicolumn{1}{|c|}{} & c1    & \multicolumn{1}{l|}{vdd}  \\ \hline
TB1                    & 100ff & \multicolumn{1}{l|}{1.2v} \\ \hline
TB2                    & 50ff  & \multicolumn{1}{l|}{1.2v} \\ \hline
TB3                    & 25ff  & 1.2v                      \\ \hline
TB4                    & 100ff & 0.8v                      \\ \hline
TB5                    & 50ff  & 0.8v                      \\ \hline
TB6                    & 25ff  & 0.8v                      \\ \hline
\end{tabular}
\end{table}
The simulation results for the proposed gates and also other well-known approaches are reported. Each of these implementations has its own merits in terms of power consumption, delay, Power-Delay-Product (PDP), area, etc., which is reported in Fig. 3-6 
 respectively. As it is shown in Fig. 3, our proposed method is better in case of area which is demonstrated from transistor count, in comparison of the other mentioned methods, except TSC\cite{self}. It can also been concluded from Fig. \ref{tbl:t2} that our method uses significantly less power in comparison of all the other mentioned methods. Hence, the proposed method shows good potential for low power applications. 
\begin{figure}[h!]
    \label{fig:area}
    \centering
    \includegraphics[width=0.5\textwidth]{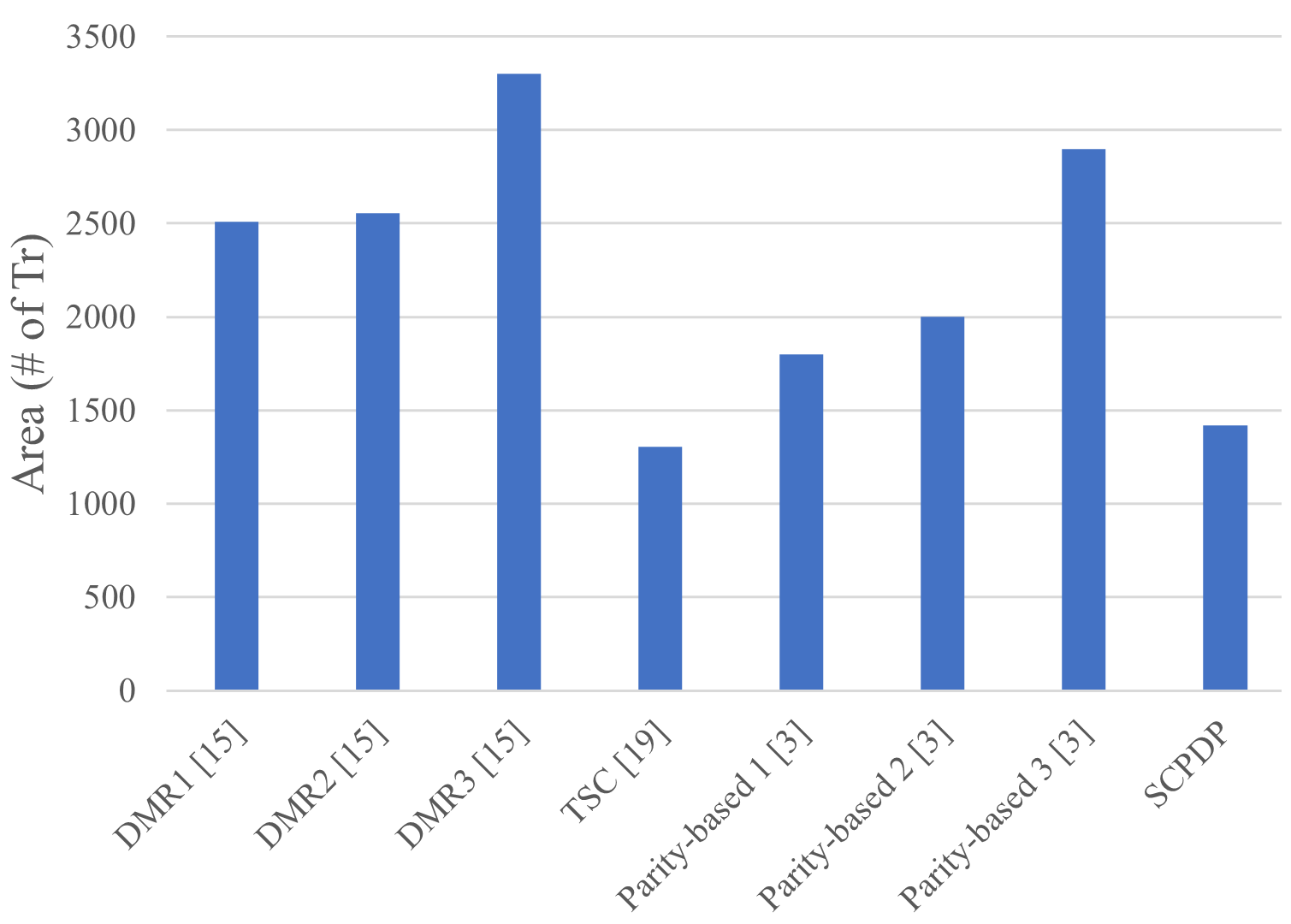}
    \caption{Technology independent area comparison of different fault-tolerant techniques on S-box.}
\end{figure}
\begin{figure*}[h!]
\centering
{\includegraphics[width=0.7\textwidth]{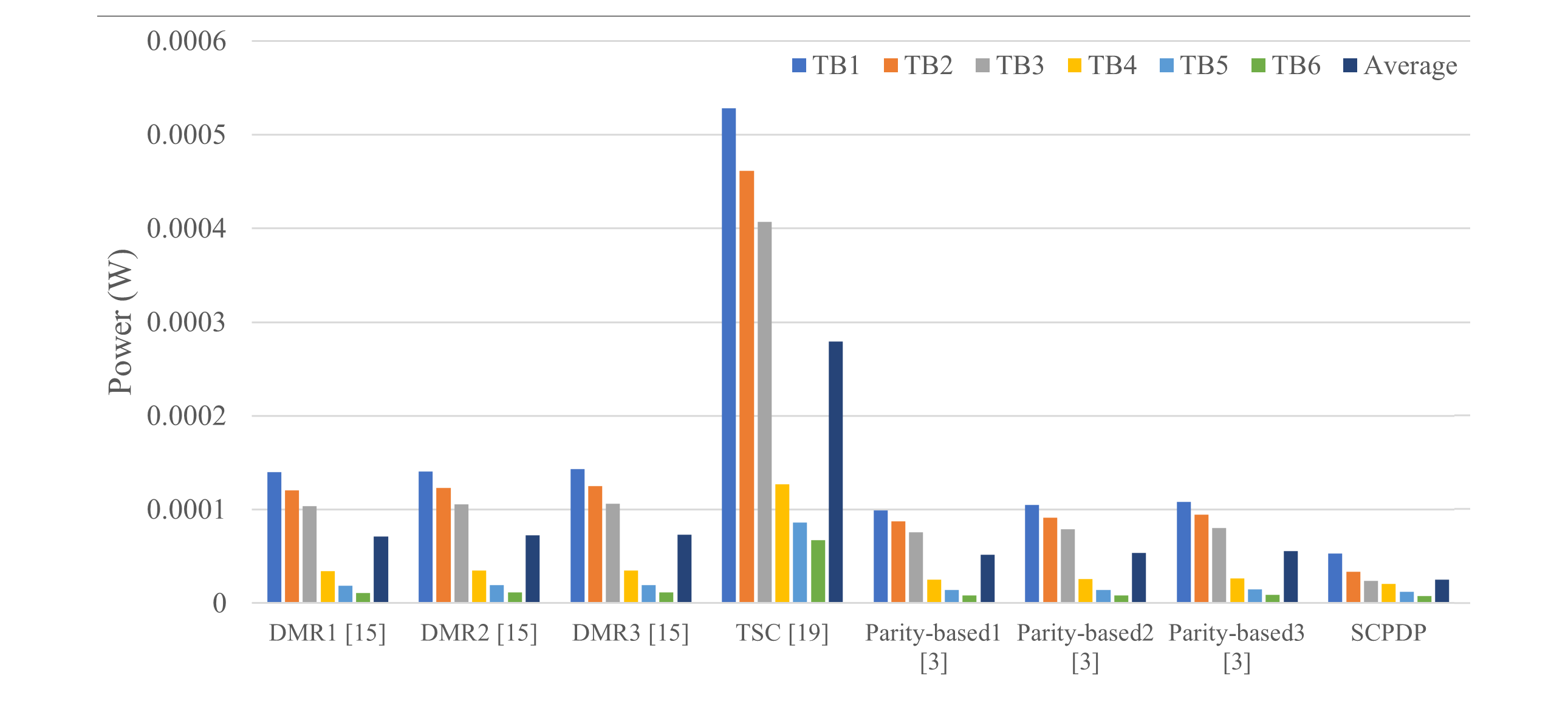}}%
\caption{Power consumption of different design.}
\label{tbl:t2}
\end{figure*}
In terms of delay, Fig. \ref{tbl:t3} is clearly demonstrating that the delay can be changed based on the different Input voltage and output capacitance combinations. It is obvious that for larger output capacitance volume, the circuit needs more time to reach a valid output and this scenario is more dramatic if a reduction in input voltage occurs. For this sake, and based on the results, it can be concluded that the TB4 has the worst delay among the other states.
The Power-Delay-Product is a well-know metric \cite{pdp} which is used widely to compare the performance of state-of-the-art methods.To measure the PDP, delay and power consumption are needed which delay has been considered as worst-case propagation delay of the circuit and internal power consumption as the average (Avg.) power consumption during a period of time (T). A trade-off between speed and power consumption can be always possible which in high-performance and low-power applications, both these terms are vital, equally. Concluding from Fig. \ref{tbl:t4}, our proposed method has the lowest PDP value among all the other methods in all of the different testing situations which shows how worthy this approach is.

As sum up, the extensive simulation results using HSPICE which are reported in this section, indicate that the case-study circuit using the proposed gates has superior speed and power compared to implementations with other error-detection capability.

%\vspace{-0.1in}

\begin{figure*}[h!]
\centering
{\includegraphics[width=0.7\textwidth]{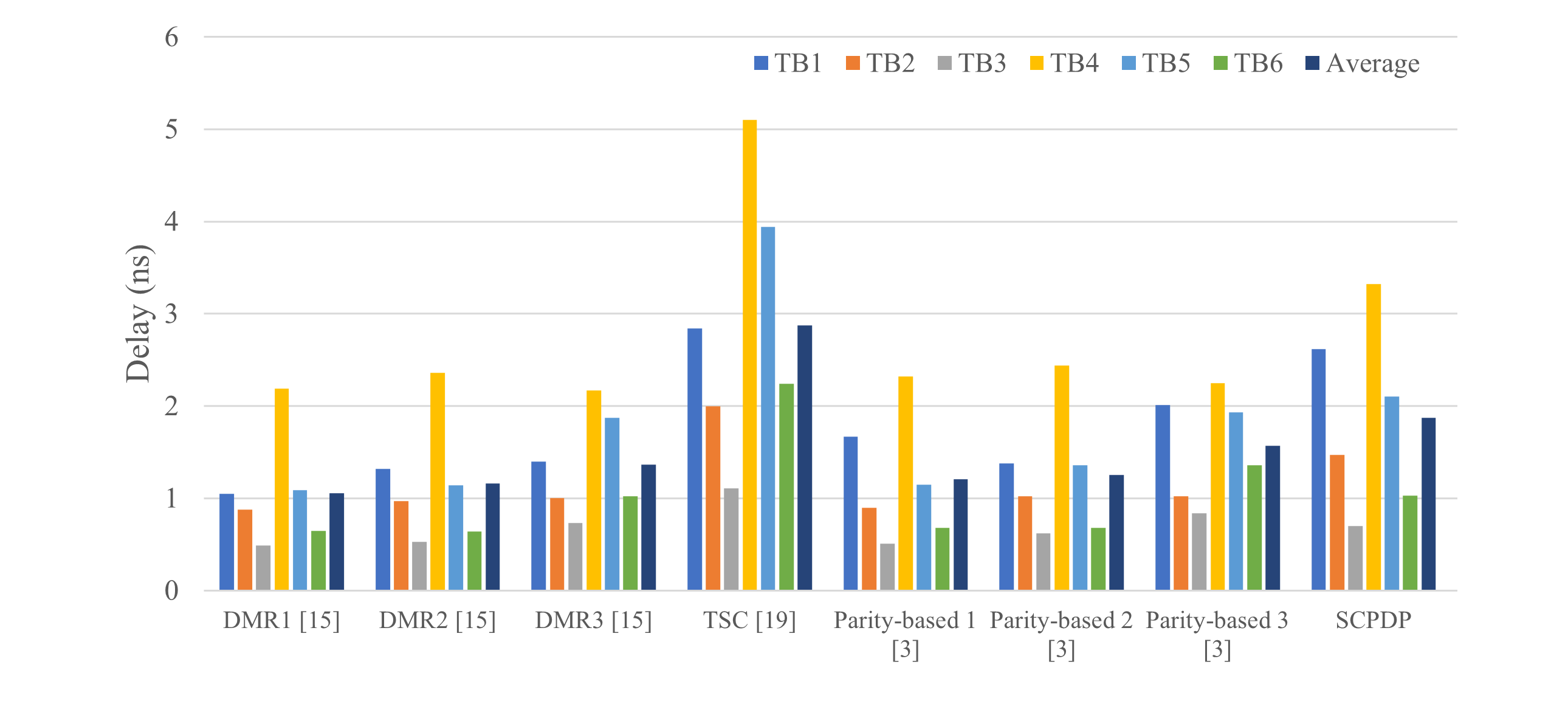}}%
\caption{Delay comparison of different fault-tolerant works on different Test-Benches (TB1-TB6).}
\label{tbl:t3}
\end{figure*}
%\begin{comment}
\begin{comment}
\begin{figure*}[h!]
\centering
{\includegraphics[width=0.7\textwidth]{figs/pdp.png}}%
\caption{Power Delay Product (PDP)  comparison. }
\label{tbl:t4}
\end{figure*}
%\vspace{-0.2in}
\end{comment}

%\end{comment}
\begin{table*}[h]
\centering
 \caption{Power-Delay product comparison}\label{tbl:t4}\label{tbl:t3}
\begin{tabular}{|l|l|l|l|l|l|l|}
\hline
PDP(nJ)           & TB1         & TB2         & TB3         & TB4         & TB5         & TB6         \\ \hline
DMRx1\cite{dmr-ref}          & 1.47E-04 & 1.06E-04 & 5.06E-05 & 7.46E-05 & 2.06E-05  & 7.14E-06 \\ \hline
DMRx2\cite{dmr-ref}           & 1.85E-04 & 1.19E-04 & 5.60E-05  & 8.17E-05 & 2.20E-05 & 7.35E-06 \\ \hline
DMRx3\cite{dmr-ref}           & 2.00E-04 & 1.24E-04 & 7.77E-05 & 7.56E-05 & 3.64E-05 & 1.20E-05  \\ \hline
TSC \cite{self}          & 1.50E-04 & 9.23E-04  & 4.54E-04 & 6.46E-04 & 3.39E-04    & 1.51E-04    \\ \hline
parity-based1 \cite{sbox0} & 1.89E-0 & 7.88E-05 & 3.87E-05  & 5.81E-05 & 1.83E-05 & 5.53E-06 \\ \hline
parity-based2 \cite{sbox0} & 1.44E-04 & 9.33E-05 & 4.88E-05 & 6.29E-05 & 1.95E-05 & 5.81E-06 \\ \hline
parity-based3 \cite{sbox0} & 2.17E-04  & 9.61E-05 & 1.92E-05 & 5.91E-05 & 2.84E-05 & 1.20E-05 \\ \hline
SCPDP (Proposed)         & 1.39E-04 & 4.95E-05 & 1.67E-05    & 5.63E-05 & 1.70E-05 & 5.32E-06 \\ \hline
\end{tabular}
\end{table*}

\section{Conclusions}
In this paper, we have proposed a novel logic style with self-checking capability, which provides excellent reliability for practical circuits implementation. The proposed logic style has been designed with a very low number of transistors, which leads to efficient hardware implementation, low power consumption and also high-performance operation.  The fault coverage for this newly proposed logic style was 99.98 $\%$ for multiple-bit faults, while the other methods achieved 97 $\%$ fault coverage at best.
We provided a detailed theoretical proof for the single-bit fault detection capability of the proposed logic style which has also been demonstrated  by extensive simulation results.
Our proposal is illustrated to be superior to the popular DMR as well as other existing fault detection structures in various design metrics, i.e. hardware efficiency and reconfigurability, power consumption and performance, making them suitable for a wide range of resource-constrained applications. 

\section*{Acknowledgments}

This work was supported in part by the European Union through European Social Fund in the frames of the ``Information and Communication Technologies (ICT) programme'' (``ITA-IoIT'' topic) and by the Estonian Research Council grant PUT PRG1467 ``CRASHLES''.
\bibliographystyle{IEEEtran}
%\bibliographystyle{cas-model2-names}


\bibliography{cas-refs}

\end{document}